\documentclass[]{spie}  
\usepackage[dvips]{graphicx}

\title{Scientific goals of the UKIRT Infrared Deep Sky Survey} 

\author{Steve Warren
\skiplinehalf
UKIDSS Survey Scientist, Astrophysics, Imperial College, London, UK
}

\authorinfo{E-mail: s.j.warren@ic.ac.uk}

\begin{document} 
  \maketitle 

\begin{abstract}
UKIDSS is the next generation near--infrared sky survey. The survey
will commence in early 2004, and over 7 years will collect 100 times
as many photons as 2MASS. UKIDSS will use the UKIRT Wide Field Camera
to survey 7500 square degrees of the northern sky, extending over both
high and low Galactic latitudes, in JHK to K=18.5 (over
three magnitudes deeper than 2MASS). UKIDSS will be the true
near--infrared counterpart to the Sloan survey, and will produce as
well a panoramic clear atlas of the Galactic plane. In fact UKIDSS is
made up of five surveys and includes two deep extra--Galactic
elements, one covering 35 square degrees to K=21, and the other
reaching K=23 over 0.77 square degrees. This paper provides the
details of the five UKIDSS surveys and describes the main science
goals.
\end{abstract}



\section{INTRODUCTION}
\label{sect:intro}  

UKIDSS is the next generation near--infrared JHK sky survey, the
successor to 2MASS. UKIDSS will use the UKIRT wide field camera WFCAM,
currently under construction in Edinburgh and scheduled for
commissioning at the end of 2003. The camera uses four Rockwell Hawaii
II $2048\times2048$ HgCdTe arrays. When commissioned WFCAM will have
the widest field of view of any near--ir camera in the world,
capturing a solid angle of 0.21 square degrees in a single exposure.

UKIDSS was the idea of Andy Lawrence (IfA, Edinburgh) who is the
UKIDSS Principal Investigator. UKIDSS in fact consists of five surveys
covering a range of depths and areas at both high and low Galactic
latitude. The UKIDSS Consortium is a collection of some 80 astronomers
who are responsible for the design and execution of the surveys. For
more details visit the website {\tt http://www.ukidss.org}. The data
become public to the whole ESO astronomy community as well as part of
the Japanese astronomy community immediately they are entered into the
archive. There will be a proprietary period, probably 18 months,
before allowing world access.

All magnitudes quoted in this article are on the Vega system.

\section{UKIRT WIDE FIELD CAMERA}

\begin{figure}
\begin{center}
\begin{tabular}{c}
\includegraphics[height=9cm]{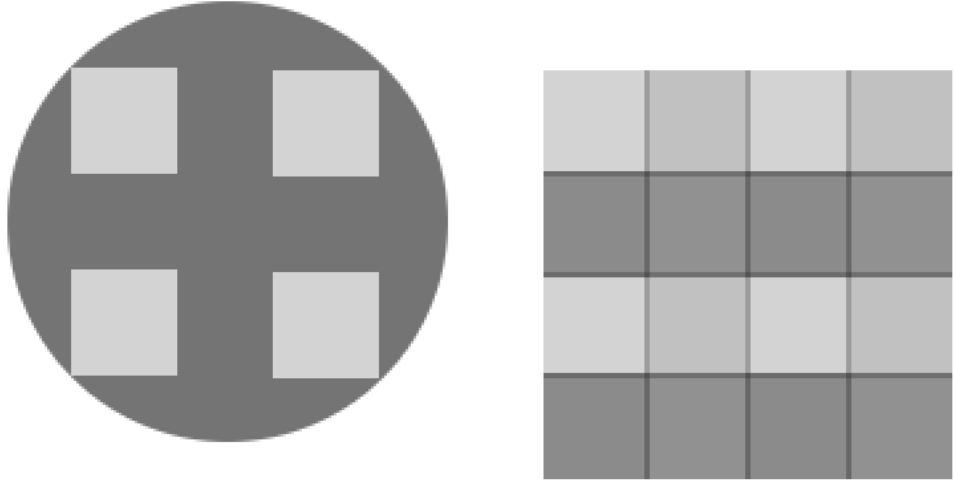}
\end{tabular}
\end{center}
\caption[example]{\label{fig:wfcam} Left: Focal plane arrangement of
WFCAM. Each array is $2048\times2048$ pixels, with pixel size 0.4
arcsec. The $2\times2$ arrangement of arrays is contained within a
circular field of diameter 0.97 degrees. Right: Filled--in tile after
$2\times2$ macro--steps. Allowing for the overlaps, the solid angle of a tile
is 0.77 square degrees. The deepest survey, the UDS, will cover a
single tile.}
\end{figure}

WFCAM is a cryogenic Schmidt--type near--ir camera under construction
at the UK ATC in Edinburgh. The focal plane will hold four
$2048\times2048$ PACE HgCdTe arrays. The spacing between detectors is
$90\%$ of the array width. The array configuration is illustrated in
Fig. \ref{fig:wfcam} LHS. The pixel size is 0.4 arcsec, so the
instantaneous exposed field of view of WFCAM is 0.21 square
degrees. With this configuration a single complete seamless $4\times4$
tile is achieved in four pointings, as illustrated in
Fig. \ref{fig:wfcam} RHS. Accounting for overlaps the solid angle of a
single tile is 0.77 square degrees.

The WFCAM focal plane has diameter one degree. The wide field of view
has been achieved by a novel forward--Cassegrain design which achieves
a high degree of off--axis correction and includes a cold pupil
stop. Further information can be found at {\tt
http://www.roe.ac.uk/atc/projects}. Tip--tilt image stabilisation will
be realised using the existing tip--tilt hexapod stage with a new $f/9$
secondary mirror. The large pixel size was chosen to maximise survey
speed. With the good image quality enjoyed by UKIRT the median seeing
psf will be undersampled. We plan to microstep with a $2\times2$,
$N+0.5$ pixel step to improve the sampling, interlacing the data to
create the final image.

\section{FIVE SURVEYS}

UKIDSS consists of five surveys exploring both high and low Galactic
latitudes to a variety of depths. The surveys will take some seven
years to complete, finishing in 2010, and will require a total of 1000
nights on UKIRT. Details of the five surveys are summarised in Table
\ref{tab:surveys}. The Large Area Survey (LAS) covers 4000 square
degrees at high Galactic latitudes, to K=18.4 ($5\sigma$). This depth is three
magnitudes deeper than 2MASS. The LAS will be the true near--infrared
counterpart to the Sloan survey. The Galactic Plane Survey (GPS) will
provide a panoramic clear atlas of the Milky Way disk, reaching
K=19.0, surveying the strip to 5 degrees above and below the plane
along a length 180 degrees. The Galactic Clusters Survey (GCS) will
undertake a fundamental study of the faint end of the stellar initial
mass function, imaging the nearest stellar clusters to a depth K=18.7.
Finally two deep surveys, the Deep Extragalactic Survey (DXS) reaching
K=21 over 35 square degrees, and the Ultra Deep Survey (UDS) reaching
K=23 over 0.77 square degrees, will study galaxies at high
redshifts. 

\begin{table}[h]
\caption{Details of the five elements of UKIDSS. The quoted depths are
total magnitude 5$\sigma$ for a point source. The estimated number of
nights for each survey includes an allowance for poor weather. The Y
band covers the wavelength range $0.97-1.07\mu$m.}
\label{tab:surveys}
\begin{center}       
\begin{tabular}{|l|c|r|c|r|} \hline
         Survey          &Filter&Area    &Mag. limit &  Nights \\
                         &      &sq. degs&  (Vega)   &         \\ \hline 
                         &  Y   &        &  20.5     &         \\
Large Area Survey        &  J   & 4000   &  20.0     &   262   \\
     LAS                 &  H   &        &  18.8     &         \\
                         &  K   &        &  18.4     &         \\ \hline
                         &  J   &        &  20.0     &         \\
Galactic Plane Survey    &  H   & 1800   &  19.1     &   186   \\
     GPS                 &  K   &        &  19.0     &         \\ \cline{2-3}
                    &$\:\:$H$_2$&  300   &  ...      &         \\ \hline
                         &  J   &        &  19.7     &         \\
Galactic Clusters Survey &  H   & 1600   &  18.8     &    84   \\
     GCS                 &  K   &        &  18.7     &         \\ \hline
                         &  J   &   35   &  22.5     &         \\
Deep Extragalactic Survey&  H   &    5   &  22.0     &   118   \\
     DXS                 &  K   &   35   &  21.0     &         \\ \hline
                         &  J   &        &  25.0     &         \\
Ultra Deep Survey        &  H   & 0.77   &  24.0     &   296   \\
     UDS                 &  K   &        &  23.0     &         \\ \hline
\end{tabular}
\end{center}
\end{table}

\begin{figure}
\begin{center}
\begin{tabular}{c}
\includegraphics[height=10cm]{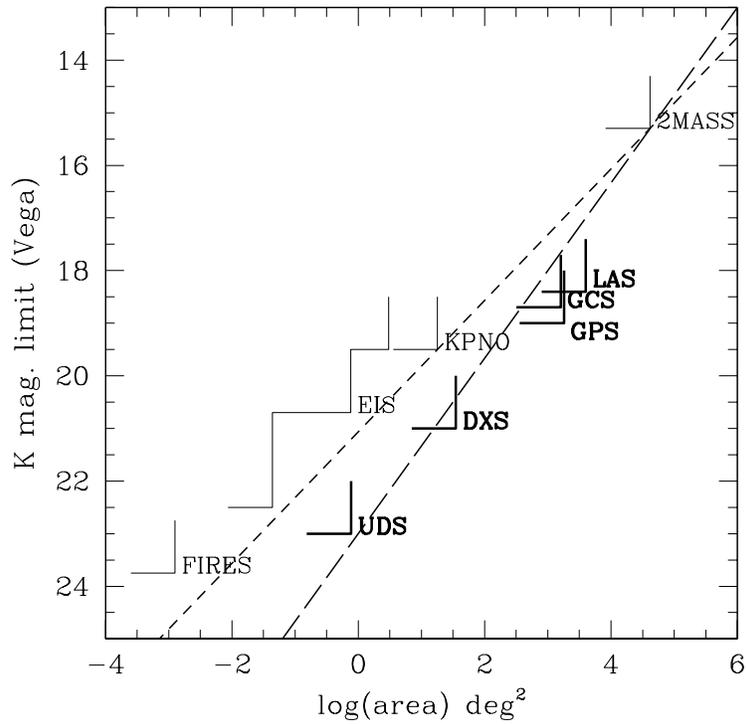}
\end{tabular}
\end{center}
\caption[example]{\label{fig:deptharea} Comparison of survey area and
$5\sigma$ depth in K of existing and planned near--ir surveys. Note
the enormous range of the X axis, 10 orders of magnitude. The most
ambitious existing or planned surveys, besides UKIDSS, are plotted
light. These are: 2MASS, the KPNO deep survey, the ESO Imaging Survey
EIS, and the very deep FIRES VLT image of HDF--S. The five UKIDSS
elements are plotted dark. The short-dashed line, normalised to 2MASS,
plots the relation $area\propto 10^{-0.8K}$, which is the trade--off
between area and depth for a fixed amount of telescope time (for
sky--limited observations). The position of a survey relative to this
line is a relative measure of the total number of photons
collected. The long--dashed line, again normalised to 2MASS, plots the
relation $area\propto 10^{-0.6K}$. The position of a survey relative
to this line is a relative measure of the volume of space
surveyed. [The different slopes of the two lines illustrate the point
that for a fixed amount of telescope time a wide, shallow survey maps
more volume than a deep, narrow survey.]}
\end{figure}

Fig. \ref{fig:deptharea} plots the combinations of depth and area of
the five elements of UKIDSS, compared against some existing or planned
near--ir surveys. The short-dashed line compares surveys against 2MASS in
terms of number of photons collected. In this sense the KPNO deep
survey and the deepest element of EIS will, when completed, be
comparable to 2MASS. Relative to this line each of the UKIDSS elements
is a factor about 30 larger in area. Taken as a whole UKIDSS is two
orders of magnitude more ambitious than any survey currently completed
or underway.

In the same figure the long-dashed line compares surveys against 2MASS
in terms of the volume mapped. This calculation assumes the spatial
distribution of objects is uniform and that space is Euclidian, so is
appropriate, for example, for estimating the relative numbers of brown
dwarfs. In this respect the summed volume of the three shallow UKIDSS
surveys, the LAS, GPS, and GCS, is about 20 times the 2MASS volume.

\begin{figure}
\begin{center}
\begin{tabular}{c}
\includegraphics[height=6.5cm]{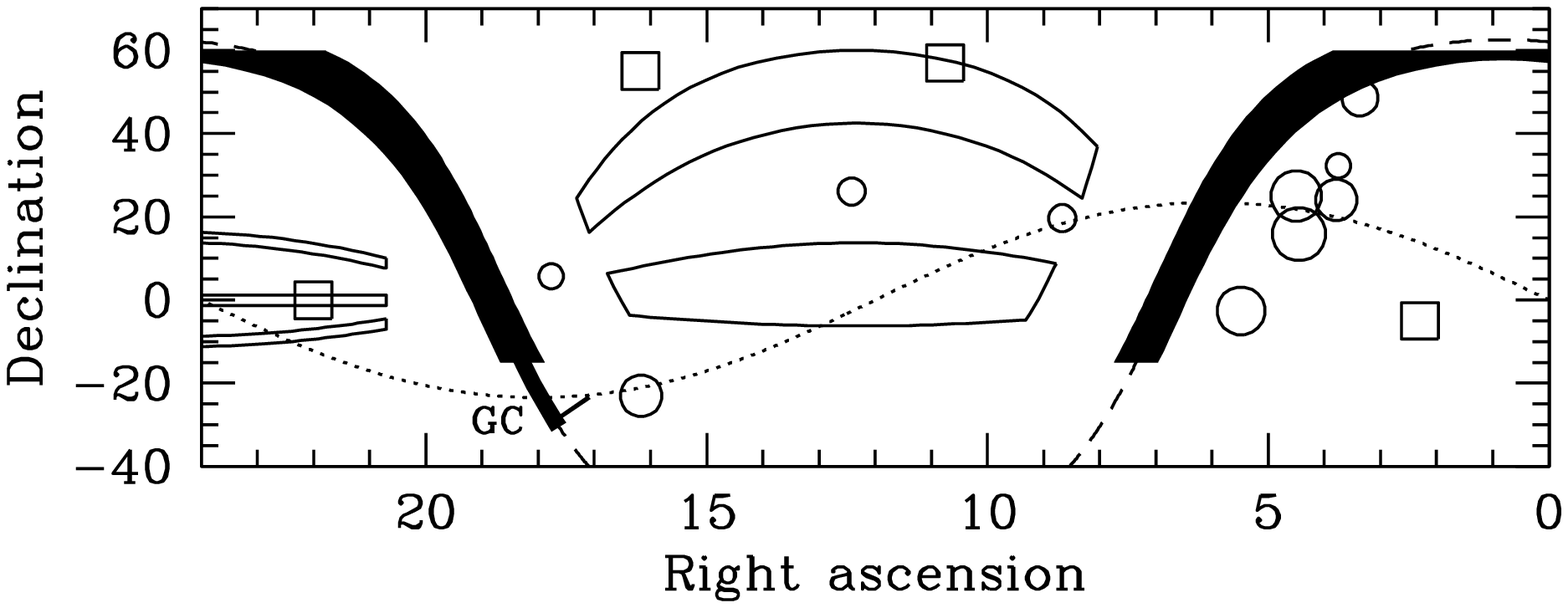}
\end{tabular}
\end{center}
\caption[example]{\label{fig:radec} This plot marks the planned
locations of the survey fields. Note that UKIRT lies at latitude
$+20^{\circ}$. The ten GCS fields are marked by circles and the four
DXS fields by squares. The UDS field is the Subaru/XMM--Newton Deep
Survey field, which lies at the centre of the DXS field located at
$J0218-05$. The GPS fields are the broad solid bands. The GPS covers
the sections of the Galactic plane within $-5^{\circ}<b<+5^{\circ}$
and $-15^{\circ}<$Dec$<+60^{\circ}$, plus a narrower extension to the
Galactic centre (marked GC). The LAS fields are the unfilled curved
bands, and are SDSS stripes. Further details are provided on the web
pages at {\tt http://www.ukidss.org}.}
\end{figure}

Fig. \ref{fig:radec} is a map showing the planned locations of the
different fields that will be covered by the five elements. After the
UK joined ESO the field selection was adjusted to a more
southerly configuration, to improve access from the VLT.

\section{SCIENCE GOALS}

\subsection{Science headlines}

The headline scientific goals of UKIDSS are the following:

\begin{itemize}
  \item to find the nearest and faintest sub--stellar objects 
  \item to break the $z=7$ quasar barrier 
  \item to determine the epoch of re--ionisation 
  \item to determine the substellar mass function 
  \item to discover Population II brown dwarfs
  \item to measure the abundance of galaxy clusters at $1<z<1.5$ 
  \item to measure the growth of structure and bias from $z=3$ to the
  present day 
  \item to determine the epoch of spheroid formation 
  \item to clarify the relationship between EROs, ULIRGs, AGN and
  protogalaxies
  \item to map the Milky Way through the dust, to several kpc 
  \item to increase the number of known Young Stellar Objects by an
  order of magnitude, including rare types such as FU Orionis stars  
\end{itemize}

The science goals of the five surveys are described individually below.

\subsection{Large Area Survey (LAS)}

The LAS was conceived as the infra--red counterpart to the Sloan
Digital Sky Survey (SDSS), as well as an atlas for identification of
sources detected in surveys at other wavelengths. SDSS fields covering
4000 square degrees will be observed in the four passbands YJHK, with
a second pass in J a few years later, for proper motions. Opening the
near--ir window will greatly clarify the make--up of low--redshift
galaxies in terms of mix of stellar populations and dust content. The
long--wavelength data also enhance the detectability of distant galaxy
clusters, as well as reddened X--ray/far--ir/radio survey
sources. However it is the opportunity for finding new classes of rare
object, because of the great volume surveyed
(Fig. \ref{fig:deptharea}), which is the most exciting prospect of the
LAS. In the search for brown dwarfs the LAS will explore an order of
magnitude more volume than 2MASS, and therefore will be the most
powerful survey for both the coolest and the least luminous dwarfs. We
hope to uncover the cool sequence extending beyond type T (called
Y dwarfs by Kirkpatrick\cite{Kirk2002}), also to detect Population II
brown dwarfs using proper motion, and to discover the lowest mass
dwarfs, possibly free--floating planets at sub--parsec distances.

\begin{figure}
\begin{center}
\begin{tabular}{c}
\includegraphics[height=7.5cm]{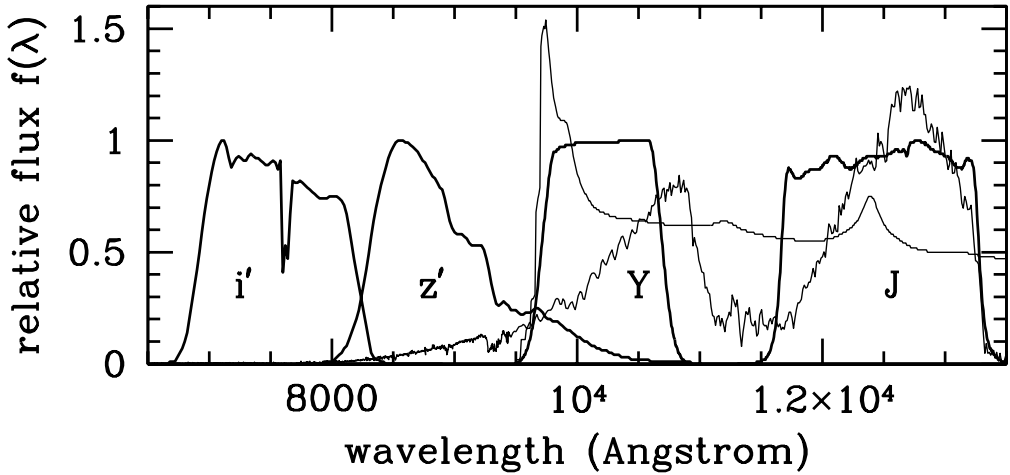}
\end{tabular}
\end{center}
\caption[example]{\label{fig:yband} This plot (from Warren and
Hewett\cite{WarrHewe2002}) shows the throughput curves of the SDSS
$i^{\prime}$ and $z^{\prime}$ filters, the standard J filter, and the
new UKIDSS Y filter, $0.97-1.07\mu$m, which will be used to detect
$z=7$ quasars and Y dwarfs. The spectrum of the T dwarf SDSS1624 (from
Leggett et al.\cite{Legg2000}) is plotted as well as a model spectrum of a
$z=7$ quasar.}
\end{figure}

We anticipate similar success in the search for the highest redshift
quasars. Building on the success of the SDSS in finding $z>6$ quasars
we hope to extend this frontier to redshift $z=7$ and beyond. This
work employs the Y band, which covers the wavelength range
$0.97-1.07\mu m$. The bandpass of Y is illustrated in
Fig. \ref{fig:yband}. This filter is distinct in wavelength from the
SDSS $z'$ band, and the $z'-Y$ colour is crucial for the highest
redshift quasars, so we have preferred to give the filter a new name
rather than call it $z(infrared)$. Hillenbrand et al.\cite{Hill2002} describe
a similar filter, and coincidentally have given it the same label.

In summary, we hope with the LAS to discover both the nearest (outside
the solar system) and the farthest known objects in the Universe.

\subsection{Galactic Plane Survey (GPS)}

The GPS will map in JHK half the Milky Way within latitude
$\pm5^{\circ}$, covering arcs of longitude $15^{\circ}<l<107^{\circ}$,
and $142^{\circ}<l<230^{\circ}$ (Fig. \ref{fig:radec}), plus a thinner
strip extending into the Galactic centre and from there up into the bulge.
The GPS will be scanned three times in K, improving the depth to
K=19.0, and providing variability information. This is deep enough to
see all the way down the IMF in distant star formation regions, to
detect luminous objects such as OB stars and post--AGB stars across
the whole Galaxy, and to detect G--M stars to several
kpc. Additionally a narrow--band molecular hydrogen survey over a
smaller area (300 square degrees) will be conducted in the
Taurus--Auriga--Perseus dark cloud region.

The principal science drivers of the GPS are: (1) creation of a legacy
database and 3--D Atlas; (2) study of star formation and the IMF with
emphasis on environmental dependence; (3) detection of counterparts to
X--ray and gamma--ray sources; (4) AGB stars, PPN and Planetary Nebulae,
including detection of brief phases of stellar evolution; and (5)
brown dwarfs: the GPS is similar in scope to the LAS in this regard.

\subsection{Galactic Clusters Survey (GCS)} 

The GCS is aimed at the crucial question of the sub--stellar mass
function. The stellar mass function is well determined down to the
brown--dwarf boundary but more or less unknown below, and the question
of whether the IMF as a whole is universal is unanswered. The GCS will
survey eleven large open star clusters and star formation associations
in JHK, with a second pass in K for proper motions. These clusters are
all relatively nearby and are several degrees across. The GCS improves
on current studies not primarily by going deeper but by collecting
much larger numbers, and examining clusters with a range of ages and
metallicities in order to address the issue of universality. The
mass limit reached varies somewhat from cluster to cluster, but is
typically near $25M_{Jupiter}$.

\subsection{Deep Extragalactic Survey (DXS)} 

The DXS will map 35 square degrees of sky to depths of K=21, and
J=22.5, in four separate regions. The four fields are XMM--LSS,
Lockman Hole, Elais--N1, and VIRMOS--4. The theme of the DXS is a
comparison of the properties of the Universe at $1.0<z<1.5$ against
the properties of the Universe today. The DXS will survey a similar
volume at these redshifts to the 2dF and Sloan Digital Sky Survey
(SDSS) volumes, and the near--infrared gives coverage of the same
rest--frame wavelengths as SDSS. Much of the DXS science relies on
multi--wavelength coverage and the choice of the four fields took
account of existing or potential coverage by XMM--Newton, GALEX,
CFHLS, VIRMOS, VST, and SIRTF.

The principal goal of the DXS, which sets the scope of the survey, is
the measurement of the abundance of rich galaxy clusters at
$1<z<1.5$. The purpose is to obtain constraints on cosmological
parameters. Ultimately we hope to make an important contribution to
reaching beyond the three--parameter $H_{\circ}$, $\Omega_m$,
$\Omega_\Lambda$ cosmology that describes the geometry and dynamics of
the Universe, to obtain useful constraints on the dark energy equation
of state parameter $w=P/\rho$, and thereby to explore quintessence
models. Two other important goals of the DXS are (i) to provide the
photometric catalogue for a redshift survey at $z>1$, similar in scope
to the 2dF galaxy redshift survey, to measure the evolution of large
scale structure, and (ii) to quantify the contribution from both star
formation and AGN to the cosmic energy budget, as a function of
wavelength over the X--ray to far--infrared region, and to measure
clustering for the different classes of object; normal galaxies (by
type), starbursts, EROs, AGN.

\subsection{Ultra Deep Survey (UDS)} 

The UDS aims to produce the first large--volume map of the Universe at
high redshift, $z=3$, surveying a region 100 Mpc comoving across and 2
Gpc deep ($2<z<4$). Essentially the aim is to go as deep as is
feasible over an area of one WFCAM tile (four pointings,
Fig. \ref{fig:wfcam}). The depth has been chosen by reference to the
detectability of a $z=3$ $L^*$ elliptical that formed at
$z=5$. Accounting for the surface--brightness profile of such a galaxy,
the required equivalent point--source depth is K=23.

The prime aim of the UDS is the measurement of the abundance of
high--redshift elliptical galaxies. Secondary goals are the
measurement of the clustering of galaxies at $z=3$, and clarification
of the relationship between EROs, ULIRGs, AGN and protogalaxies. The
abundance of ellipticals at high redshift is a key test of
hierarchical theories of structure formation, and the scope of the UDS
is sufficient to distinguish between current competing models. The
inspiration for the UDS came from observing the success of the HDF as
a public legacy database. In the same way we expect that the UDS will
be used for many important science projects that have not yet been
thought of.

\acknowledgments
The science case presented here was developed by the UKIDSS
consortium. The UKIDSS PI is Andy Lawrence. The Survey Heads are
Richard Jameson (LAS), Phil Lucas (GPS), Nige Hambly (GCS), Alastair
Edge (DXS), and Omar Almaini (UDS).

\bibliography{spieukidss}   
\bibliographystyle{spiebib}   

\end{document}